\documentclass[10pt, conference, letterpaper]{IEEEtran}
\IEEEoverridecommandlockouts
\usepackage{cite}
\usepackage{amsmath,amssymb,amsfonts}
\usepackage{algorithmic}

\usepackage{graphicx}
\graphicspath{{./}}
\usepackage{subcaption}

\usepackage{textcomp}
\usepackage{xcolor}

\usepackage{balance}

\usepackage[english]{babel}
\usepackage{hyphenat}

\usepackage[dvipdfm,bookmarksopen=true,colorlinks,citecolor=black,linkcolor=black,anchorcolor=black,urlcolor=black] {hyperref}

\newtheorem{theorem}{Theorem}
\newtheorem{claim}{Claim}

\def\BibTeX{{\rm B\kern-.05em{\sc i\kern-.025em b}\kern-.08em
    T\kern-.1667em\lower.7ex\hbox{E}\kern-.125emX}}

\begin{document}

\title{When Load Rebalancing Does Not Work for Distributed Hash Table
}

\author{\IEEEauthorblockN{Yuqing Zhu}
\IEEEauthorblockA{zhuyuqing@tsinghua.edu.cn}
}

\maketitle

\begin{abstract}
Distributed hash table (DHT) is the foundation of many widely used storage systems, for its prominent features of high scalability and load balancing. Recently, DHT-based systems have been deployed for the Internet-of-Things (IoT) application scenarios. Unfortunately, such systems can experience a breakdown in the scale-out and load rebalancing process. This phenomenon contradicts with the common conception of DHT systems, especially about its scalability and load balancing features. In this paper, we investigate the breakdown of DHT-based systems in the scale-out process. We formulate the load rebalancing problem of DHT by considering the impacts of write workloads and data movement. We show that, the average network bandwidth of each node and the intensity of the average write workload are the two key factors that determine the feasibility of DHT load rebalancing. We theoretically prove that load rebalancing is not feasible for a large DHT system under heavy write workloads in a node-by-node scale-out process.
\end{abstract}


\section{Introduction}%
\label{sec:intro}

Distributed hash table (DHT)~\cite{dht} is a storage infrastructure providing a lookup service similar to a hash table. It is the foundation  underlying modern large-scale key-value stores, e.g., Dynamo~\cite{dynamo} and Cassandra~\cite{cassandra}, which are basic building blocks for large-scale production systems in the cloud. DHT-based systems can easily scale out to a larger size when the application workload increases. Composed of networked nodes, DHT-based systems can balance application workloads across different nodes, guaranteeing assured experiences for applications.

DHT provides measures for guaranteeing the two key features of scalability and load balancing for cloud-based systems. It distributes keys to a set of system nodes through consistent hashing~\cite{consistentHashing}. Values are stored along with their keys at the corresponding system nodes. As for load balancing, the hashing can distribute the keys almost uniformly across system nodes, given the assumption that the number of keys is large. With regard to scalability, data access can be handled by any system node without any centralized bottleneck; and, the system can scale out by adding more nodes.

\emph{\bf Load rebalancing} of DHTs happens on system stabilization, which is a process for preserving desirable properties after system changes, e.g., system scaling out by adding more nodes. On system scaling out, key ranges and values are moved across system nodes to preserve the mapping relations between keys and nodes. By such, system load gets rebalanced. In general cases, the system scale-out events are not often and can be planned by system administrators.

Previous works~\cite{p2pevolv,chord} on DHT only consider how key routing and node membership are affected during system stabilization. The impacts of data movement for load rebalancing has hardly been studied. When the write workload for the system is relatively low as in the past, the number of nodes is the main factor affecting system stabilization and the routing complexity is the major concern. However, the impact of load rebalancing on the system cannot be ignored when the write workload of incoming data is heavy.

With the advent of Internet-of-Things (IoT), there emerges a need for supporting the workload of unprecedentedly frequent yet steady writes. Previously, a system will handle millions of requests a day~\cite{dynamo}. Or, for the biggest Cassandra cluster in the world~\cite{biggestCCluster}, a system can handle millions of reads and writes per second, using tens of thousands of nodes. As for the IoT applications, DHT-based systems remain a popular choice~\cite{kairosdb,opentsdb}, but now the system has to handle millions of writes per second over only tens of nodes. A write workload that is too heavy can cause the breakdown of a DHT-based system. The following is a related use case for storing IoT data in the DHT-based system Cassandra.

\noindent\textbf{Example - Metro IoTs.} \emph{A city has 20 metro lines. Each line is served by about 10 trains. Each train is equipped with 3200 sensors to monitor its operating status. Each sensor is sending back a value larger than 16 bytes at a 2Hz frequency. Besides monitoring the train status on the recent data, the metro company wants to keep all the data for future analytics tasks. To fulfill the task, the storage system must handle a workload of {\small $20\times10\times3200\times2=1,200,000$} writes per second.}\\
\emph{A Cassandra-based system named KairosDB~\cite{kairosdb} is exploited for the purpose. The storage system is composed of 14 nodes initially, with each node serving about $85,000$ writes per second. Afterwards, the sampling frequency of each sensor is raised from 2Hz to 5Hz. Besides, the number of trains for each lines is increased to 15 to enable a better public transport service. Therefore, the storage system must now handle {\small $20\times15\times3200\times5=4,800,000$} writes per second. The storage system exhibits incompetency in the process, which makes the system administrators decide to scale out the system to increase the system capacity.}\\
\emph{The planned scale-out process is composed of a sequence of node addition procedures. Each procedure adds one node to the system. The next procedure will not begin until the system gets stabilized in the previous procedure. Writes continue to arrive at the system during the scale-out process. Unfortunately, the Cassandra storage system breaks down at the beginning of the planned scale-out event, when the first node is added and when the system rebalances loads. After a careful analysis, we find out that the breakdown does not happen due to any bug of the Cassandra system. Rather, it is an inherent problem of DHT that lies at the core of Cassandra. This problem only exposes its traits when the write workload exceeds a certain level and during load rebalancing.}

Regarding IoT application scenarios and workload changes as in the above example, we formulate the load rebalancing problem of DHTs and address the problem from a theoretical perspective in this paper. We investigate how the intensity of writes affects system load rebalancing. We study the conditions when load rebalancing is feasible for DHT. \emph{Not until the recent wide adoptions of IoTs does the design assumptions of DHT change.} Therefore, we analyze the changes of assumptions in practice and the factors that determine the feasibility of the load rebalancing process. Our results show that \emph{DHT cannot guarantee load rebalancing and scalability with its original design, while simultaneously supporting application scenarios where writes are extremely intensive}, e.g., storing IoT data. We make the following contributions in this paper:
\begin{itemize}
  \item We formulate the load rebalancing problem of DHT by considering the impacts of write workloads and data movement in the course (\emph{Section~\ref{sec:problem}}), targeting at application scenarios emerging recently (e.g., IoT).
  \item We show that, the average network bandwidth of each node and the intensity of the average write workload are the two key factors that determine the feasibility of DHT load rebalancing. For four varied application scenarios (\emph{Section~\ref{sec:liwrite}-\ref{sec:noswrite}}), we prove theoretical bounds on the maximum rates of writes that a DHT-based system can handle, if the system is to rebalance loads in a node-by-node scale-out process.
  \item We apply the theoretical results on load rebalancing for analyzing the real case of IoT (\emph{Section~\ref{sec:app}}).
\end{itemize}

\section{Background and Related Work}

In this section, we first describe the concept and a typical design of a distributed hash table (DHT). Then, we explain the load distribution mechanisms commonly applied to DHT-based systems in practice. Finally, we position our work in the background of related works.

\subsection{Distributed Hash Table}

Distributed hash table (DHT) designs emerge along with peer-to-peer (P2P) systems such as Gnutella~\cite{gnutella} and Freenet~\cite{freenet} at the beginning of this century. DHT designs feature a structured key-based routing, enabling a decentralized organization of nodes and a guaranteed key-matching time. Typical DHT designs include Chord~\cite{chord}, CAN~\cite{can}, and Pastry~\cite{pastry}. The Chord design has been adopted in two widely deployed systems, i.e., Amazon's Dynamo~\cite{dynamo} and Apache Cassandra~\cite{cassandra}, which have supported a variety of production systems in the cloud. In the following, we assume the Chord design for DHT, but the results in this paper can be easily extend to other DHT designs with minor changes.

DHT uses consistent hashing~\cite{consistentHashing} to spread $K$ keys uniformly over $N$ connected nodes. Each node is assigned $T$ tokens. The tokens are mapped to the wrapped one-dimensional key space such that each token corresponds with a key range. Each node keeps routing information about $O(\log N)$ other nodes and resolves the mapping by communicating with $O(\log N)$ nodes. Each node only needs to coordinate with only a few other nodes in the system such that only a limited amount of work needs to be done for each change in node membership. Therefore, DHT-based system should function efficiently even with more than thousands of nodes.

To meet the requirement of cloud applications, the cloud key-value (KV) store~\cite{dynamo} makes slight modifications to the design of DHT. The modifications are to achieve better load balance, which is highly desirable for a cloud-based system. First, a system node does not only keep the routing table for its responsible keys, but also the data objects corresponding to the keys. With a balanced distribution of keys, the data will also be distributed evenly across nodes. Second, on system changes, e.g., node joining or leaving, data objects must move along with the corresponding keys, the responsibility of which changes hands. This will preserve the previous balanced load after the system changes. Third, the mapping between keys and tokens is adapted to ensure uniform load distribution on load rebalancing.

\subsection{Load Rebalancing and Uniform Distribution}

Load rebalancing can happen when the system scales out by adding nodes or scales in by removing nodes. We focus our discussions in this paper on the system scale-out event. The results can be applied to the scale-in process with slight adaptations as well, since the two processes are the reverse of each other. We consider the system scale-out event triggered on the condition that the data volume at each node has reached a threshold. Once the system starts to scale out, it faces an unstable time when nodes must migrate data and exchange routing information across the system. We denote this process as a \emph{stabilization} process, as the system seeks to stabilize for preserving the desirable system properties. System stabilization begins when the node set changes and the data migrate. It ends when the routing information and the data on each node match the system design. Then, the system can have uniform load distribution.

Three common strategies for load distribution exist~\cite{dynamo}. They vary mainly on how tokens are assigned to nodes and how key ranges are partitioned. According the different choices on the two aspects, we denote the three strategies as \emph{limited-token-random-part}, \emph{limited-token-equal-part}, and \emph{many-token-equal-part}. The \emph{limited-token-random-part} strategy assigns $T$ random tokens per node and partitions key ranges by token values. The \emph{limited-token-equal-part} strategy assigns $T$ random tokens per node and split the key ranges into a large number of equal partitions. The \emph{many-token-equal-part} strategy splits the key ranges into $Q$ equal partitions and assign $Q/N$ random tokens to each of the $N$ nodes. The first strategy is the original key-token mapping strategy of DHT, thus it can enable an extremely high scalability. But among the three strategies, the third performs the best with regard to load balance and the size of metadata~\cite{dynamo}, thus it is widely used in the cloud.

For the \emph{many-token-equal-part} strategy, the identifier circle is divided into $Q$ equal sized partitions/ranges. The right bound of each partition is taken as a token. Each node is then randomly assigned with $Q/N$  tokens, i.e., $Q=TN$. Thus, $Q$ need not be large and the metadata size is smaller than for the second strategy. With a replication factor $r$, a partition is mapped to the first $r$ unique nodes that are encountered when walking the circle clockwise from the wrapping position. On joining, a node picks one token from the other $N$ nodes. On departure, a node hands its tokens randomly to the remaining nodes. Our discussion assumes this strategy in this paper.

\subsection{Related Work}

Chord~\cite{chord}, CAN~\cite{can} and Pastry~\cite{pastry} are among the best known distributed hash table (DHT) systems. They are efficient for exact key matches, guaranteeing a $O(\log N)$ search length. Their mapping between keys and nodes enables a localized computation of routing tables and an effective adaptation under system churns, i.e., changes in node membership~\cite{churn}. Therefore, systems based on such DHTs can scale to an extremely large size.

DHT has been used as the basis for a variety of systems, e.g., key-value stores~\cite{hyperdex} and file systems~\cite{chordfs}. Amazon Dynamo~\cite{dynamo} and Apache Cassandra~\cite{cassandra} are among the best known systems that are based on DHTs. These two systems are key-value (KV) stores. They work as the foundations of many cloud systems, thanks to their prominent property of scalability and load balancing. While DHT enables the KV stores to scale easily, these systems adapt the key-node mapping strategies to achieve load balance. In this paper, our discussions of DHT assumes the load balancing property.

Extensive research efforts have been devoted to analyze different aspects of DHT designs, e.g., load balancing~\cite{dhtloadbalance}, the routing cost~\cite{routing}, churn handling~\cite{chrunAndRnumber}, and malicious attacks~\cite{attacks}. Load \emph{rebalancing} has also been studied~\cite{dhtreloadbalance}. However, previous works mainly study these topics with regard to \emph{churns} and key-search workloads. \textbf{The impact of write workload on the system has rarely been taken into account.} Similarly, while works exist on analyzing load balance of services composed of distributed KV stores~\cite{kvloadbalance}, the write workloads are not considered in the studies. The reasons are that, DHTs are mainly used for indexing objects previously, while key-value stores have yet to meet a heavy write workload that can actually make an impact, until the emergence of IoT data.

Our work differs from previous works in that we analyze the load rebalancing property of DHT with regard to a potential heavy workload of writes, which come from the new application scenario of IoT. The results of our work should apply to the wide application scenarios of IoT and offer implications on the scalable storage system design for IoT data.
\begin{table}[t]
\caption{Symbols used throughout the paper.}\vspace{-6pt}
\begin{center}
\begin{tabular}{|l|l|}
\hline
\multicolumn{1}{|c|}{\textbf{Symbols}} &\multicolumn{1}{|c|}{\textbf{Uses}} \\
\hline
$N$ & Node number\\
\hline
$T$ & Token number per node\\
\hline
$Q$ & Key space partition number\\
\hline
$K$ & Key number\\
\hline
$v$ & Mean size of a key-value pair\\
\hline
$r$ & Replication factor\\
\hline
$\lambda$ & Arrival rate of writes per second per node\\
\hline
$b$ & Bandwidth of a node\\
\hline
\end{tabular}\vspace{-12pt}
\label{tbl:symbol}
\end{center}
\end{table}
\section{Problem Formulation}
\label{sec:problem}

We first introduce the DHT system model along with some important early results. We then specify the load rebalancing problem with regard to the system expansion process and the stabilization procedure. The symbols used throughout the paper are summarized in Table~\ref{tbl:symbol}.

\subsection{System Model}

A DHT-based system consisted of $N$ nodes is initially in a stable state, storing $K$ keys and the corresponding values. After the node join process, the system will remain balanced. The process between balanced states is called stabilization. According to the paper introducing consistent hashing~\cite{consistentHashing}, the following results exist for the system.
\begin{theorem}
With $K$ keys hashed to $N$ nodes, the following is true with high probability.
\begin{enumerate}
  \item Each node is responsible for at most $(1+\epsilon)\frac{K}{N}$ keys.
  \item When an $(N+1)$th node joins or leaves the system, the responsibility for $O(\frac{K}{N})$ key changes hands (and only to or from this joining/leaving node).
\end{enumerate}
\end{theorem}

The consistent hashing paper also shows that $\epsilon$ can be reduced to arbitrarily small by having each node run $\Omega(logN)$ virtual nodes, each of which has also a token. As a result, the $K$ keys in the system can be distributed across $N$ nodes almost evenly, if each physical node is to host multiple virtual nodes.

Based on the above theorem, when adopting the many-token-equal-part load distribution strategy, each node serves $\frac{K}{N}$ keys along with the corresponding $\frac{K}{N}$ data values. Assuming the mean size of each value is $v$, data take up about $v\frac{K}{N}$ space at each node. With a replication factor $r$, each node serves $r\frac{K}{N}$ keys and has at least $rv\frac{K}{N}$ storage space.

\subsection{System Expansion and Stabilization}
\label{sec:expansion}

The system initiates an expansion procedure on a certain condition. An example condition can be when the ingestion workload is too heavy to be supported by the current number of nodes. Each system expansion procedure involves only the join process of a single node. To add more nodes, we can break it down to multiple one-node join process. Assuming the system has steady write workloads before expansion. The average write rate for each node is $\lambda$. At the time $T_{N+1}$ when node $N+1$ joins, there are $K$ keys and values in the system. We have:
\begin{equation}
K=\lambda NT_{N+1}\label{eq:keywrite0}
\end{equation}

\emph{Migrated keys.} The system stabilization procedure occurs during the system expansion process to guarantee the load rebalancing. On system stabilization, data must be moved from the original $N$ nodes to the joining node. With the \emph{many-token-equal-part} strategy, the joining node will receive $k^{N+1}$ keys from the other $N$ nodes. Assuming homogeneous structure, i.e., that the $N$ nodes are symmetric, each node will hand $k_i^{N+1}$ keys over to the new node. Based on the system model, we have the following equations.
\begin{align}
k^{N+1}=\mathop{\Sigma}_i^N k_i^{N+1}&=\frac{rK}{N+1}\label{eq:key0}\\
k_i^{N+1}&=\frac{rK}{N(N+1)}\label{eq:key1}
\end{align}

\emph{Data movement.} The joining node will receive $d^{N+1}$ from the other $N$ nodes, each of which sends $d_i^{N+1}$ data. Based on Eq.~\eqref{eq:key0} and \eqref{eq:key1}, we have the following relations.
\begin{align}
d^{N+1}=\mathop{\Sigma}_i^N d_i^{N+1}&=\frac{rvK}{N+1}\label{eq:data0}\\
d_i^{N+1}&=\frac{rvK}{N(N+1)}\label{eq:data1}
\end{align}

\emph{Storage space and data volume.} Assuming each node has storage space $S$. On the $N+1$ node joining, each node has used $\mu S$ storage space. Thus, each node will move $d_i^{N+1}$ data to the joining node, and the joining node receives from other nodes the following amount of data in the process:
\begin{align}
d^{N+1}=Nd_i^{N+1}=\frac{\mu SN}{N+1}\label{eq:data5}\\
d_i^{N+1}=\frac{\mu S}{N+1}\label{eq:data4}
\end{align}
Along with Eq.~\eqref{eq:data0} and \eqref{eq:data5}, we can deduce the relations between the system storage space and the number of keys as follows.
\begin{equation}
{\bf K}{\bf =\frac{\mu SN}{rv}}\label{eq:data7}
\end{equation}

\subsection{Problem Specification}

We address the load rebalancing problem of DHT-based systems. As for the example use case in Section~\ref{sec:intro}, it is obviously the load rebalancing procedure that has caused the system to break down during its expansion. However, it is unknown which factors have caused the breakdown. Two possible reasons for the system breakdown are:
\begin{enumerate}
  \item The system workload has increased during the expansion process, thus overloading and breaking down the system.
  \item The system stabilization process is receiving writes during load rebalancing, thus overloading and breaking down the system.
\end{enumerate}

With regard to the above two reasons, we need to address two aspects for the load rebalancing problem. The first aspect is about write workloads. We consider two kinds of write workloads, i.e., one with writes linearly increasing along with the system size and one with writes remaining stable during the system expansion. The second aspect is about the stabilization process. We consider both system stabilization that receives writes in the course and that does not. A stabilization that does not receive writes simultaneously is denoted as \emph{clear stabilization} in the following. For clear stabilization, writes will accumulate in the course. They are then processed after the stabilization. In the following, we investigate the load rebalancing problem with regard to four different application scenarios, which are the combinations of the above two aspects. We aim to answer the following questions:
\begin{itemize}
  \item Which factors play a key role in the load rebalancing process of a DHT-based system?
  \item What constraints exist for these factors?
\end{itemize}

Our study mainly considers the write workloads. We have simplified the problem and disregarded the impacts of the read workloads on load rebalancing. However, as we are analyzing the constraints of storage, bandwidth and time on load rebalancing, our results of bounds still hold even with the read workloads. Adding the impacts of the read workloads would only lead to tighter bounds in the results.

\section{Stabilization with Linearly Increasing Writes}%
\label{sec:liwrite}

\emph{Target application scenario:} A system can expand as the write workload increases. A node is added to the system when the write workload increases from $N\lambda$ to $(N+1)\lambda$. During the stabilization process, the system rebalances loads by migrating data across nodes. Writes continue to arrive and get processed in the process.

\emph{Results:} We can deduce two bounds for the load rebalancing process, with regard to storage constraint and bandwidth constraint respectively.

\begin{theorem}[A Storage-Oriented Bound for Rebalancing with Increasing Writes]
For a DHT-based symmetric system with $N$ nodes, each node has bandwidth $b$. The system adds one node, when each original system node has stored data up to a ratio $\mu$ of its total storage space. Assuming each write is about $v$ bytes on average. To guarantee a successful load rebalancing process, the following inequation must be satisfied for the average rate $\lambda$ of writes per node:
\begin{equation}
\lambda<(1-\frac{N}{N+1}\mu)\frac{b}{v}\label{eq:lbound6}\vspace{-6pt}
\end{equation}
\label{thm:ww:sob}
\end{theorem}

\begin{theorem}[A Bandwidth-Oriented Bound for Rebalancing with Increasing Writes]
For a DHT-based symmetric system with $N$ nodes, each node has bandwidth $b$. The system adds one node for expansion. Assuming each write is about $v$ bytes on average.  To guarantee a successful load rebalancing process, the following inequation must be satisfied for the average rate $\lambda$ of writes per node:
\begin{equation}
\lambda<\frac{1}{(N+1)}\frac{b}{v}\label{eq:tbound3}\vspace{-6pt}
\end{equation}
\label{thm:ww:bob}
\end{theorem}

Though with different degrees of bounds, the above two theorems indicate the same fact. That is, \textbf{the rate of writes that a DHT-based system can handle stays the same as the system size increases}, if the system accepts an increased workload of writes during load rebalancing.

The results seem to contradict with our previous conception of distributed key-value stores, which have high scalability disregard of the system size. The major difference lies in the assumption. Under the IoT application scenario, the workload of the system can increase as the new node is added to the system. In comparison, previous applications scenarios apply a light and relatively stable workload to the system. The workload on the system generally does not increase obviously during the expansion process.

\subsection{Preliminary}

To prove the above theorems, we first assume the joining process take time $t^{N+1}$. The joining node has to process $\lambda t^{N+1}$ writes during the joining process. The total volume of data flushing to the joining node in the course consists of two parts, i.e., the incoming writes and data moved from the other $N$ nodes. Therefore, considering Eq.~\eqref{eq:key0}, the joining node has to handle the following volume of data:
\begin{equation}
D^{N+1}=\frac{rvK}{N+1}+v\lambda t^{N+1}\label{eq:data2}
\end{equation}

Assuming that each node has enough bandwidth to transfer data to the joining node. Let $b_{N+1}$ be the part of network bandwidth that the joining node uses for the joining process.  Except for the bandwidth used for receiving writes, the joining node is assumed to use all the rest bandwidth for the data movement from other nodes. Along with Eq.~\eqref{eq:data2}, we have:
\begin{align}
b-b_{N+1}&=v\lambda\label{eq:bandwidth0}\\
t^{N+1}&=\frac{rvK}{(N+1)b_{N+1}}\label{eq:jtime1}
\end{align}

\subsection{Proof Sketch: A Storage-Oriented Bound}

To prove Theorem~\ref{thm:ww:sob}, we start from the following storage constraint on the system:
\begin{claim}
Writes should not fill the remaining storage space up at any node during the joining time for the new node.
\label{claim:storage}
\end{claim}
In other words, the time to fill up the rest $(1-\mu)S$ space at each node should be larger than the joining time $t^{N+1}$. Considering Eq.~\eqref{eq:jtime1}, the following relation between processing times is required:
\begin{equation}
\frac{(1-\mu)S}{v\lambda}>t^{N+1}\label{eq:lbound0}
\end{equation}

Filling Eq.~\eqref{eq:data7} into Eq.~\eqref{eq:lbound0}, we have the following:
\begin{equation}
(\mu^{-1}-1)(1+\frac{1}{N})b_{N+1}>v\lambda\label{eq:lbound2}
\end{equation}
Replacing $v\lambda$ in the above equation based on Eq.~\eqref{eq:bandwidth0}, we get the required relation between $b_{N+1}$ and $b$ as follows:
\begin{align}
\frac{b_{N+1}}{b}&>\frac{\mu N}{N+1-\mu}\label{eq:lbound3}\\
\Rightarrow b_{N+1}>\frac{N}{N+1-\mu}\mu b&>\frac{N}{N+1}\mu b\label{eq:lbound4}
\end{align}

We can also replace $b_{N+1}$ of Eq.~\eqref{eq:lbound3} based on Eq.~\eqref{eq:bandwidth0}. Thus, the following inequation between the rate of writes at each node $\lambda$ and the bandwidth per node exists:
\begin{equation}
\lambda<\frac{(N+1)}{N}\frac{(1-\mu)b}{v}\label{eq:lbound5}
\end{equation}
And, by replacing $b_{N+1}$ of Eq.~\eqref{eq:lbound4} based on Eq.~\eqref{eq:bandwidth0}, we can further deduce Eq.~\eqref{eq:lbound6}, and thus prove Theorem~\ref{thm:ww:sob}.

\subsection{Proof Sketch: A Bandwidth-Oriented Bound}

Assuming the system will add a new node automatically when the triggering condition is met. The condition is defined such that \emph{the system must expand when the system node has $\mu S$ data locally}. To prove Theorem~\ref{thm:ww:bob}, we start from the following bandwidth constraint on the system:
\begin{claim}
The rate for data migration at any of the $N$ node is higher than the rate of its data arrival in the course.
\label{claim:bandwidth}
\end{claim}

If the above claim is not met, a new system expansion will be triggered before the previous expansion finishes. With consecutive expansion overlapping, the system will never get stabilized or be able to serve according to its design. As each node is still having the write workload of $v\lambda$, the following inequation must be satisfied to guarantee the claim:
\begin{equation}
\frac{\mu S}{N+1}\frac{1}{t^{N+1}}>v\lambda\label{eq:tbound0}
\end{equation}

Replacing $t^{N+1}$ and $S$ in Eq~\eqref{eq:tbound0} based on Eq.~\eqref{eq:data7}, \eqref{eq:bandwidth0} and \eqref{eq:jtime1}, we can deduce the following inequations:
\begin{equation}
b>(N+1)v\lambda\label{eq:tbound2}
\end{equation}
With Eq.~\eqref{eq:tbound2}, we have thus proved Theorem~\ref{thm:ww:bob}.

\section{Clear Stabilization\\with Linearly Increasing Writes}%
\label{sec:noliwrite}

\emph{Target application scenario:} The application scenario is similar to that in Section~\ref{sec:liwrite}, except for the following difference. The system stops receiving any write until the stabilization finishes, though writes will accumulate and must be processed by the system after the stabilization process.

\emph{Results:} We can deduce a bound for the load rebalancing process based on the condition whether the accumulated writes can be processed completely between two expansion processes.

\begin{theorem}[A Time-Oriented Bound for Clear Rebalancing with Increasing Writes]
For a DHT-based symmetric system with $N$ nodes, each node has bandwidth $b$. The system adds one node, when each original system node has stored data up to a ratio $\mu$ of its total storage space. Assuming each write is about $v$ bytes on average. To guarantee a successful load rebalancing process without the interference from a write workload, the following inequation must be satisfied for the average rate $\lambda$ of writes per node:
\begin{equation}
\lambda<\frac{\sqrt{4N+1}-1}{2N}\frac{b}{v}\label{eq:nowrite14}
\end{equation}
\label{thm:nww:tob}
\end{theorem}

According to the above theorem, the write workload should not be too heavy. The indication is two fold. First, if the write workload is too heavy, e.g., as heavy as taking up almost the total bandwidth of each node, a system should not adopt the clear stabilization. Otherwise, the stabilization process would be too long to be practical. Second, if clear stabilization is adopted and the write workload is almost proportional to the system size, the system size cannot be too large. Otherwise, either the system expansion cannot complete, or the incoming writes cannot be completely received and processed.

\subsection{Preliminary}

Let the joining node process a total amount $D^{N+1}$ of data in the course of system stabilization. As writes are stopped on system expansion, we can deduce the following equalities:
\begin{align}
D^{N+1}=d^{N+1}\label{eq:nowrite1}\\
b=b_{N+1}\label{eq:nowrite2}
\end{align}

Now, the time for the stabilization process is solely determined by how fast data migration finishes. Based on Eq.~\eqref{eq:data5}, Eq.~\eqref{eq:nowrite1} and \eqref{eq:nowrite2}, we can compute the time for stabilization as follows.
\begin{equation}
t^{N+1}=\frac{d^{N+1}}{b}=\frac{\mu SN}{(N+1)b}\label{eq:nowrite3}
\end{equation}

Assuming that $N$ is sufficiently large at the beginning, we have $\frac{N}{N+1}\rightarrow 1$. $\mu S$ is the volume of data that a node stores before the system expansion. According to the above equation, \textbf{if the write workload is as heavy as to take up almost the total bandwidth $b$ of each node, when adding one more node to the system, it would take the same time to stabilize as to fill the system up before the stabilization}. In other words, Eq.~\eqref{eq:nowrite3} actually refuses the feasibility of scaling out a large system under an extremely heavy workload.

Now, we assume that the average write workload only takes up about $\alpha b$ bandwidth of each node. At the end of the system stabilization period, the accumulated data can be computed based on the stabilization time and the used bandwidth:
\begin{equation}
d_{acc}=t^{N+1}\times(N+1)\alpha b=\mu SN\alpha\label{eq:nowrite4}
\end{equation}

After the stabilization process, the bandwidth to be used for catching up the data storage would be $(1-\alpha)b$, since the rest $\alpha b$ is used for accepting new writes. Therefore, it will take the following time to catch up:
\begin{equation}
t_{catchup}=\frac{d_{acc}}{(N+1)(1-\alpha)b}=\frac{\mu SN \alpha}{(N+1)(1-\alpha)b}\label{eq:nowrite5}
\end{equation}
According to the above equation, we can observe that, \textbf{as the system size increases, the time taken for catching up is also increasing}.

After a clear stabilization, the system has $\mu SN$ data and $(N+1)S$ storage space. Before the next expansion, the system can store $\mu S$ more data. $T_{N+1}$ is the time from the system's initial status till the first expansion is initiated with one node added. And, $T_{N+2}$ is the time when the next expansion happens. We can compute $T_{N+1}$, as well as the time between this first expansion and the next one in the following:
\begin{align}
T_{N+1}&=\frac{\mu S}{\alpha b}\label{eq:nowrite7}\\
T_{N+2}-T_{N+1}&=\frac{\mu S}{(N+1)\alpha b}\label{eq:nowrite6}
\end{align}

\subsection{Proof Sketch: A Time-Oriented Bound}

To prove Theorem~\ref{thm:nww:tob}, we start from the following time constraint on the stabilization process:
\begin{claim}
The time for catching up must finish before the next system expansion event happens.
\label{claim:time}
\end{claim}
According to the above claim, with Eq.~\eqref{eq:nowrite6} and \eqref{eq:nowrite5}, the following inequality exists:
\begin{align}
T_{N+2}-T_{N+1}&>t_{catchup}\label{eq:nowrite8}\\
\Rightarrow\frac{1}{N}&>\frac{\alpha^2}{1-\alpha}\label{eq:nowrite10}
\end{align}

According to the result of Eq.~\eqref{eq:nowrite10}, if clear stabilization is adopted under a workload with linearly increasing writes, \textbf{there exist a moment when $N$ grows so large that the system expansion cannot catch up with the write workload}.

Consider the relation between the incoming write workload and the corresponding bandwidth taken up at each node, we have $v\lambda=\alpha b$. Replacing all terms of $\alpha$ in Eq.~\eqref{eq:nowrite10} accordingly, we have the following result:
\begin{align}
\lambda^2+\frac{b}{Nv}\lambda&<\frac{b^2}{Nv^2}\label{eq:nowrite12}\\
\Rightarrow (\lambda+\frac{b}{2Nv})^2&<\frac{4Nb^2+b^2}{4N^2v^2}\label{eq:nowrite13}\\
\Rightarrow \lambda&<\frac{\sqrt{4N+1}-1}{2N}\frac{b}{v}\label{eq:nowrite15}
\end{align}
Solving the inequation~\eqref{eq:nowrite13} with regard to $\lambda$, we reach the result of Eq.\eqref{eq:nowrite15} and thus Theorem~\ref{thm:nww:tob}.

\section{Stabilization with Stable Writes}%
\label{sec:swrite}

\emph{Target application scenario:} A system can expand when the write workload increases. A node is added to the system for an expansion. The write workload on the system is kept stable before and after the expansion process. On system stabilization, the system rebalances loads by migrating data across nodes. Writes continue to arrive and get processed in the process.

\emph{Results:} We can deduce two bounds for the load rebalancing process, with regard to storage constraint and bandwidth constraint respectively.

\begin{theorem}[A Storage-Oriented Bound for Rebalancing with Stable Writes]
For a DHT-based symmetric system with $N$ nodes, each node has bandwidth $b$. The system adds one node, when each original system node has stored data up to a ratio $\mu$ of its total storage space. Assuming each write is about $v$ bytes on average. To guarantee a successful load rebalancing process, the following inequation must be satisfied for the stable write workload $\lambda N$ on the system:
\begin{equation}
\lambda<(1+\frac{1}{N}-\mu)\frac{b}{v}\label{eq:slbound7}\vspace{-6pt}
\end{equation}
\label{thm:wsw:sob}
\end{theorem}

\begin{theorem}[A Bandwidth-Oriented Bound for Rebalancing with Stable Writes]
For a DHT-based symmetric system with $N$ nodes, each node has bandwidth $b$. The system adds one node for expansion. Assuming each write is about $v$ bytes on average.  To guarantee a successful load rebalancing process, the following inequation must be satisfied for the stable write workload $\lambda N$ on the system:
\begin{equation}
\lambda<\frac{1}{N}\frac{b}{v}\label{eq:stbound3}\vspace{-6pt}
\end{equation}
\label{thm:wsw:bob}
\end{theorem}

Comparing the results of Theorem~\ref{thm:ww:sob} and \ref{thm:wsw:sob}, we find that, for the two application scenarios, the bounds of $\lambda$ differ at a value of $\delta=\frac{1}{N}-\frac{\mu}{N+1}$. As $\delta>0$, the bound in Eq.~\eqref{eq:slbound7} is larger than that in Eq.~\eqref{eq:lbound6}. Thus, for the application scenario of this section, the system allows a higher rate $\lambda$ of writes per node. Theorem~\ref{thm:ww:bob} and \ref{thm:wsw:bob} also indicate the same result.

Furthermore, according to Theorem~\ref{thm:ww:bob} and \ref{thm:wsw:bob}, the bound for the write workload is related not only to the bandwidth, but also with the system size. \textbf{This correlation between $\lambda$ and $N$ explains the cause of scalability problems in the DHT-based systems that have load rebalancing processes}. When $\lambda$ is small as for previous application scenarios, the bound is too far to reach and the problem would not exist. However, for the IoT application scenarios where continuous and highly frequent writes come from machines, the bound can be reached easily. Hence, scalability problems occur in DHT-based storage systems in practice, as in the use case of Section~\ref{sec:intro}.

Finally, based on Theorem~\ref{thm:wsw:bob}, it is required that \textbf{the global write workload $N\lambda$ for the system must be smaller than the maximum writes that a single node can receive, i.e., $\frac{b}{v}$}. Otherwise, the system will have scalability problems, if its expansion process involves load reblancing.

\subsection{Preliminary}

To prove the above theorems, we first assume the joining process take time $t^{N+1}$. Writes arrive at the rate of $N\lambda$ per second at the system. As the system load is assumed to be balanced, the joining node needs to process $\frac{N}{N+1}\lambda t^{N+1}$ writes on joining. Along with the rebalanced data as in Eq.~\eqref{eq:key0}, the joining node must process the following amount of data the course of system stabilization:
\begin{equation}
D^{N+1}=\frac{rvK}{N+1}+\frac{N}{N+1}\lambda vt^{N+1}\label{eq:sdata1}
\end{equation}

Assuming that each node has enough bandwidth to transfer data to the joining node. Let $b_{N+1}$ be the part of network bandwidth that the joining node uses for the joining process. Except for the bandwidth $b_{N+1}$ used for load rebalancing, the joining node uses all the rest bandwidth for receiving writes, which are about $\frac{1}{N+1}$ of the $\lambda N$ workload. Considering Eq.~\eqref{eq:sdata1}, we can deduce the following relations:
\begin{align}
b-b_{N+1}&=\frac{N}{N+1}\lambda v\label{eq:sdata2}\\
t^{N+1}&=\frac{rvK}{(N+1)b_{N+1}}\label{eq:sdata3}
\end{align}

\subsection{Proof Sketch: A Storage-Oriented Bound}

Again, we start from the storage constraint on the system, i.e., Claim~\ref{claim:storage}, for proving Theorem~\ref{thm:wsw:sob}. Considering Eq.~\eqref{eq:sdata3}, the following relation between processing times is required:
\begin{equation}
\frac{(N+1)(1-\mu)S}{N\lambda v}>t^{N+1}\label{eq:slbound0}
\end{equation}

To replace $t^{N+1}$ and $S$, we fill in Eq.~\eqref{eq:slbound0} with Eq.~\eqref{eq:data7} and \eqref{eq:sdata3}. Hence, the following inequality is deduced:
\begin{align}
(\mu^{-1}-1)(1+\frac{1}{N})b_{N+1}&>\frac{N}{N+1}\lambda v\label{eq:slbound2}
\end{align}

We first deduce the relation between $b_{N+1}$ and $b$. We remove the terms of $\lambda$ exploiting Eq.~\eqref{eq:sdata2}. The following results exist:
\begin{align}
b_{N+1}&>\frac{N}{N+1-\mu}\mu b\label{eq:slbound3}\\
\Rightarrow b_{N+1}&>\frac{N}{N+1}\mu b\label{eq:slbound4}
\end{align}

Now, we fill in Eq.~\eqref{eq:slbound4} with Eq.~\eqref{eq:sdata2} to replace the terms of $b_{N+1}$. By such, we obtain the result of Eq.~\eqref{eq:slbound7}, proving Theorem~\ref{thm:wsw:sob}.

\subsection{Proof Sketch: A Bandwidth-Oriented Bound}

Again, we assume the system will expand and add a new node when each node of the system has $\mu S$ data on average. The system has stable writes. Hence, while each node has a write workload of $\lambda$ before expansion, its workload is reduced to $\frac{N}{N+1}\lambda$ after the new node joins.

To prove Theorem~\ref{thm:wsw:bob}, we start from Claim~\ref{claim:bandwidth}, which must be guaranteed by the system. Considering Eq.~\eqref{eq:data4}, we can deduce the following relation:
\begin{equation}
\frac{\mu S}{N+1}\frac{1}{t^{N+1}}>\frac{N}{N+1}\lambda v\label{eq:stbound0}
\end{equation}
According to Eq.~\eqref{eq:stbound0}, the rate of writes at any node must be smaller than the data transfer rate during the system stabilization process. Otherwise, the storage limit $\mu S$ will be exceeded and Claim~\ref{claim:bandwidth} cannot be guaranteed.

Replacing $t^{N+1}$ and $S$ in Eq~\eqref{eq:stbound0} based on Eq.~\eqref{eq:data7}, \eqref{eq:sdata2}, and \eqref{eq:sdata3}, we can deduce the following inequalities:
\begin{align}
b_{N+1}&>\frac{N^2}{N+1}\lambda v\label{eq:stbound1}\\
\Rightarrow b&>N\lambda v\label{eq:stbound2}
\end{align}
With Eq.~\eqref{eq:stbound2}, we have thus proved Theorem~\ref{thm:wsw:bob}.

\section{Clear Stabilization on Stable Writes}%
\label{sec:noswrite}

\emph{Target application scenario:} The application scenario is similar to that in Section~\ref{sec:swrite}, except for the following difference. The system stops receiving any write until the stabilization finishes, though writes will accumulate and must be processed by the system after the stabilization process.

\emph{Results:} We can deduce a bound for the load rebalancing process based on the condition whether the accumulated writes can be processed completely between two expansion processes.

\begin{theorem}[A Time-Oriented Bound for Clear Rebalancing with Stable Writes]
For a DHT-based symmetric system with $N$ nodes, each node has bandwidth $b$. The system adds one node, when each original system node has stored data up to a ratio $\mu$ of its total storage space. Assuming each write is about $v$ bytes on average. To guarantee a successful load rebalancing process without the interference from a write workload, the following inequation must be satisfied for the stable write workload $\lambda N$ on the system:
\begin{equation}
\lambda<\frac{(N+1)(\sqrt{4N+1}-1)}{2N^2}\frac{b}{v}\label{eq:noswrite16}
\end{equation}
\label{thm:nwsw:tob}
\end{theorem}

\vspace{-12pt}With the above theorem, we can make a similar observation as in Section~\ref{sec:noliwrite} that the system size cannot grow too large for a heavy write workload. This observation is valid regardless of whether the write workload increases or remains stable after system expansion. In other words, \textbf{when the write workload is heavy enough, a DHT-based system will come into the scalability problem such that the load cannot be rebalanced on system stabilization}.\vspace{-6pt}

\subsection{Preliminary}

Let the joining node process a total amount $D^{N+1}$ of data in the course of system stabilization. As writes are stopped on system expansion, we can deduce the following equalities:
\begin{align}
D^{N+1}=d^{N+1}\label{eq:noswrite1}\\
b=b_{N+1}\label{eq:noswrite2}
\end{align}

Filling Eq.~\eqref{eq:data5} in Eq.~\eqref{eq:noswrite1}, we can deduce the time for stabilization based on Eq.~\eqref{eq:noswrite2} as follows:
\begin{equation}
t^{N+1}=\frac{d^{N+1}}{b}=\frac{\mu SN}{(N+1)b}\label{eq:noswrite3}
\end{equation}
Notice that, Eq~\eqref{eq:noswrite3} is the same as Eq.~\eqref{eq:nowrite3}. Hence, \textbf{a large system under an extremely heavy workload cannot scale out by adding one node by a clear stabilization process}, no matter the write workload has increased or keeps stable in the course. Here, a large system means a system with a large size $N$.

Under a clear stabilization, writes will accumulate to be processed later, thus requiring a catching-up process afterwards. Assuming that the average write workload takes up about $\alpha b$ bandwidth of each node before system expansion. When the stabilization process completes, the accumulated data can be computed based on Eq.~\eqref{eq:noswrite3}:
\begin{equation}
d_{acc}=t^{N+1}\times N\alpha b=\frac{N^2}{N+1}\mu S\alpha\label{eq:noswrite4}
\end{equation}

After the stabilization process, the bandwidth to be used for catching up the data storage would be $b'_{N+1}$ at each node. As the pre-expansion workload of writes is $N\alpha b$ and the workload is stable, each node will need to spend $\frac{N}{N+1}\alpha$ of its bandwidth for processing writes. Thus, we have the following relation between $b'_{N+1}$ and $b$:
 \begin{equation}
b'_{N+1}=(1-\frac{N}{N+1}\alpha)b\label{eq:noswrite5}
\end{equation}

Combining Eq.~\eqref{eq:noswrite4} with \eqref{eq:noswrite5}, we can deduce the catching-up time as follows:
\begin{equation}
\footnotesize
t_{catchup}=\frac{d_{acc}}{(N+1)b'_{N+1}}=\frac{N^2}{(N+1)(N+1-N\alpha)b}\mu S\alpha\label{eq:noswrite6}
\end{equation}

As described in Section~\ref{sec:expansion}, the system has $\mu SN$ data before expansion. After a clear stabilization, the volume of data does not change, but the total storage space increases from $NS$ to $(N+1)S$. Assuming that an expansion is initiated only when the data volume has reached the limit. On the next expansion, each system node will have $\mu S$ data. Hence, the system can store $\mu S$ more data before the next expansion.

$T_{N+1}$ is the time from the system's initial status till the first expansion is initiated with one node added. And, $T_{N+2}$ is the time when the next expansion happens. Since the write workload remain stable, the system will have a write workload of $N\alpha b$ before and after the expansion. Now, we can compute $T_{N+1}$, as well as the time between this first expansion and the next one as follows:
\begin{align}
T_{N+2}-T_{N+1}=\frac{\mu S}{N\alpha b}\label{eq:noswrite8}\\
T_{N+1}=\frac{\mu S}{\alpha b}\label{eq:noswrite9}
\end{align}

Assuming that the write workload remain unchanged for a sufficiently long time. Based on Eq.~\eqref{eq:noswrite8}, we can accordingly deduce the time between any two consecutive expansion, which is the same as $T_{N+2}-T_{N+1}$. Therefore, if a DHT-based system can scale out and rebalance loads in the course, with a stable write workload, the system must expand at a stable speed, disregarding the size $N$ of the system.
\begin{figure*}[t]
    \begin{subfigure}{0.33\textwidth}%
    \centering%
        \includegraphics[width=\textwidth]{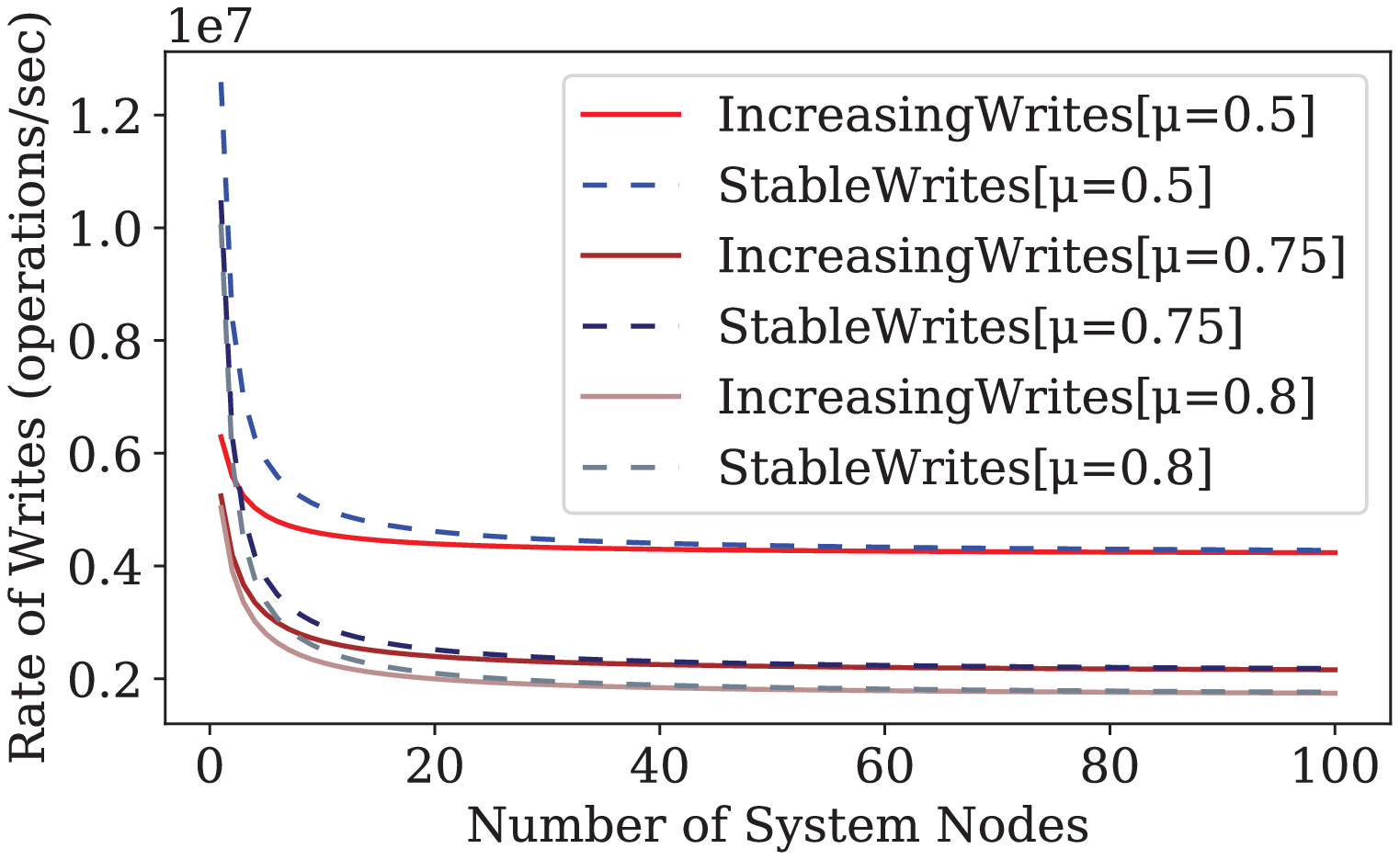}%
        \caption{Storage-Oriented Bounds}
        \label{fig:result:sob} 
    \end{subfigure}
    \begin{subfigure}{0.33\textwidth}%
    \centering%
        \includegraphics[width=\textwidth]{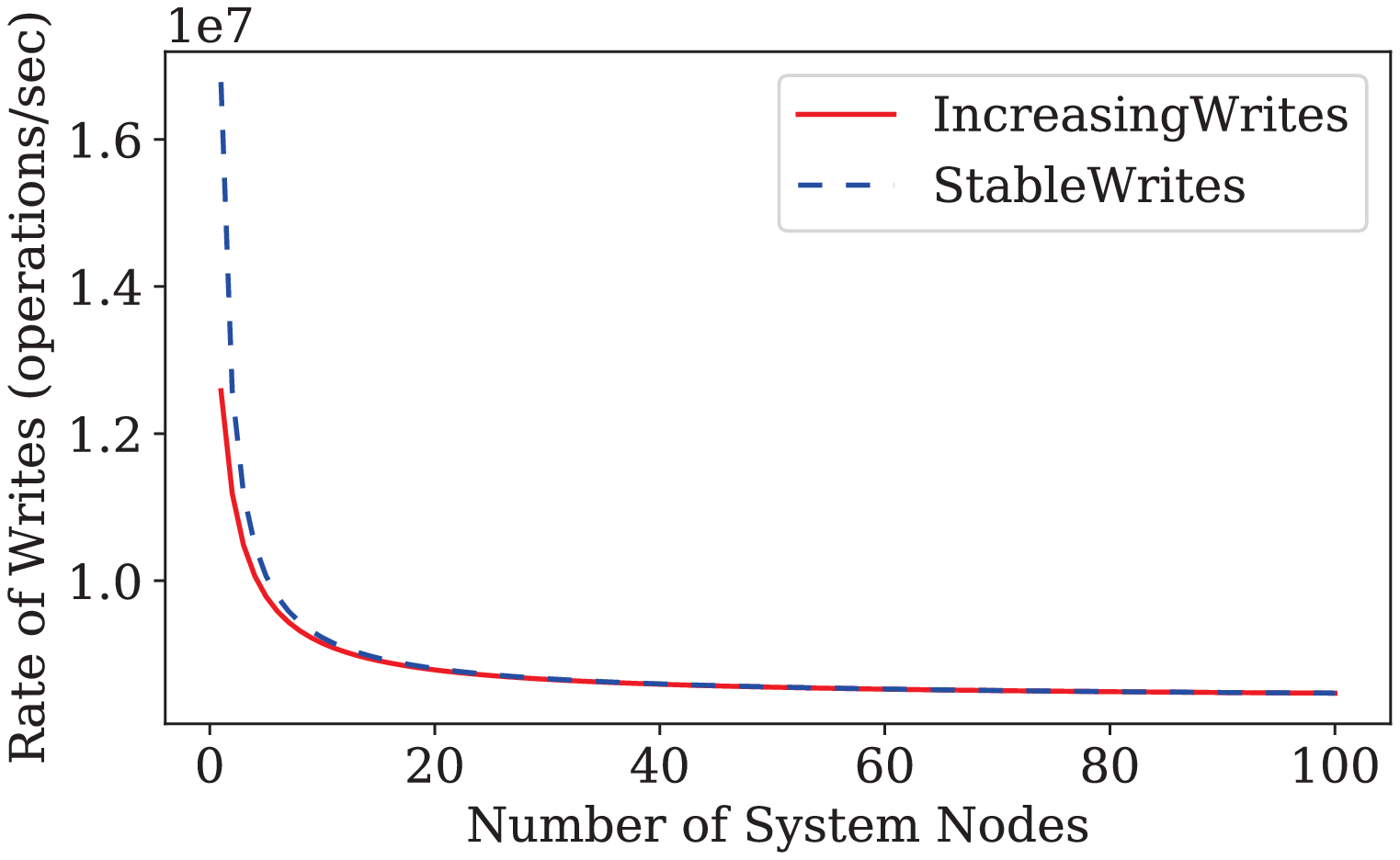}%
        \caption{Bandwidth-Oriented Bounds}
        \label{fig:result:bob} 
    \end{subfigure}
    \begin{subfigure}{0.33\textwidth}%
    \centering%
        \includegraphics[width=\textwidth]{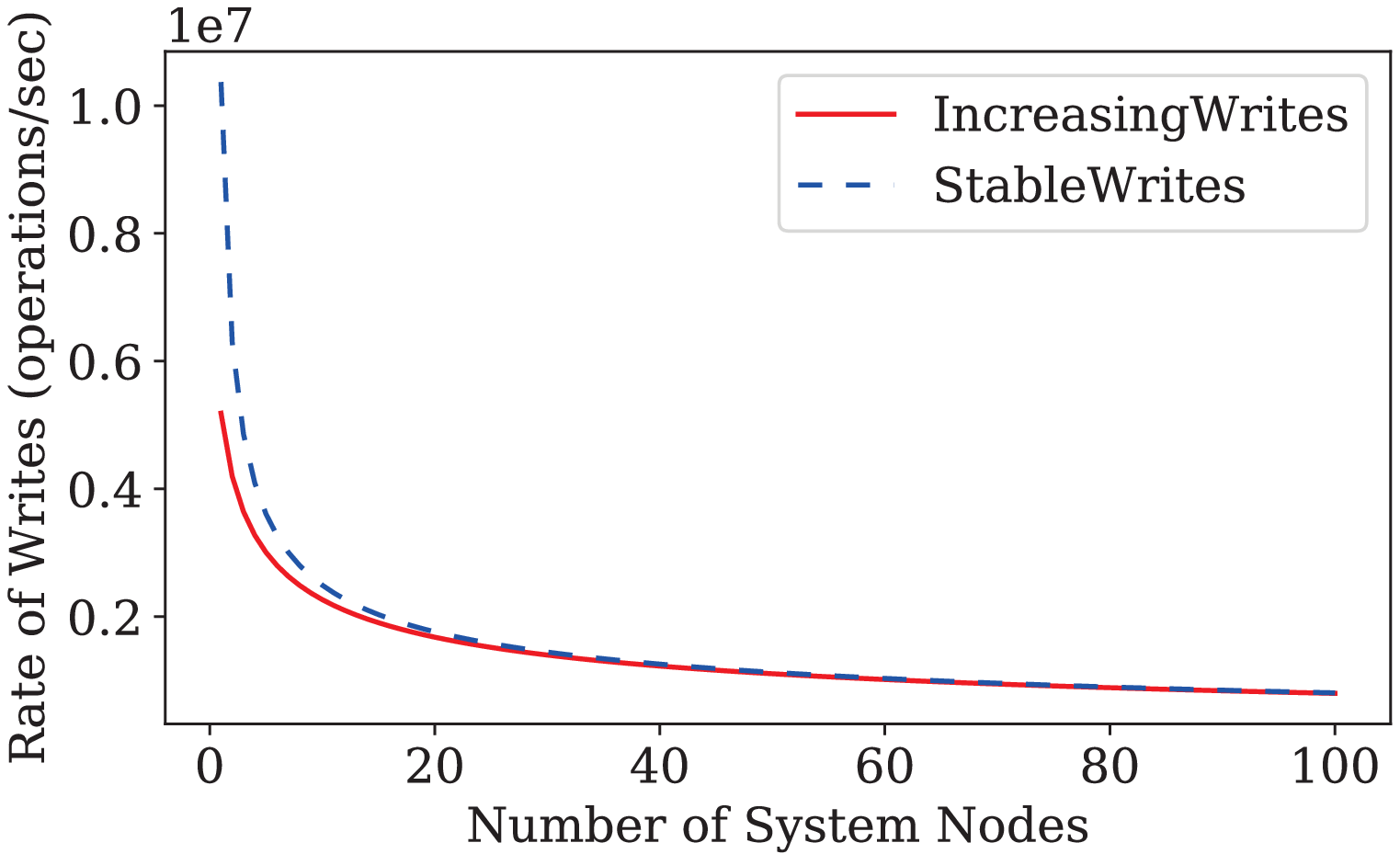}%
        \caption{Time-Oriented Bounds}
        \label{fig:result:tob} 
    \end{subfigure}
    \caption{The relation curves between the rate of writes and the system size $N$, according to Theorem[\ref{thm:ww:sob},\ref{thm:wsw:sob}], Theorem[\ref{thm:ww:bob},\ref{thm:wsw:bob}], and Theorem[\ref{thm:nww:tob},\ref{thm:nwsw:tob}].}
    \label{fig:result} 
\end{figure*}
\subsection{Proof Sketch: A Time-Oriented Bound}

We now sketch the proof for Theorem~\ref{thm:nwsw:tob}. Starting from Claim~\ref{claim:time}, we obtain the following relations:
\begin{equation}
T_{N+2}-T_{N+1}>t_{catchup}\label{eq:noswrite10}
\end{equation}
Filling in Eq.~\eqref{eq:noswrite10} with Eq.~\eqref{eq:noswrite6} and \eqref{eq:noswrite8}, we can deduce the following result:
\begin{equation}
\alpha^2+\frac{N+1}{N^2}\alpha<\frac{(N+1)^2}{N^3}\label{eq:noswrite12}
\end{equation}

According to the result of Eq.~\eqref{eq:noswrite12}, we have $N(\frac{N^2}{N+1}\alpha^2+\alpha-1)<1$. Since $1-\alpha$ is smaller than 1, one can notice that $N$ cannot be too large with regard to a specific $\alpha\in(0,1)$. Thus, if clear stabilization is adopted under a workload with stable writes, \textbf{the system size $N$ cannot grow too large, if load rebalancing is demanded during system stabilization}.

For each node, the arrived writes and the consumed bandwidth are equal before the system expansion, i.e., $v\lambda=\alpha b$. Along with \eqref{eq:noswrite12}, we deduce the following inequation:
\begin{align}
(\lambda+\frac{(N+1)b}{2N^2v})^2&<\frac{b^2(N+1)^2(4N+1)}{4N^4v^2}\label{eq:noswrite15}\\
\Rightarrow \lambda&<\frac{(N+1)(\sqrt{4N+1}-1)}{2N^2}\frac{b}{v}\label{eq:noswrite17}
\end{align}
Solving the above inequation~\eqref{eq:noswrite15} with regard to $\lambda$, we obtain the result of Eq.~\eqref{eq:noswrite17} and thus \eqref{eq:noswrite16}, getting Theorem~\ref{thm:nwsw:tob} proved.

\section{Applications}
\label{sec:app}%
In this section, we demonstrate the practical value of our theoretical results by first examining the corresponding numerical results in real-world settings. Then, we apply these results for analyzing an IoT application scenario.\vspace{-6pt}

\subsection{Numerical Results}
\label{sec:numerical}

\textbf{Storage-Oriented Bounds.} We start with the storage-oriented bounds for stabilization with linearly increasing writes and with stable writes. The related theoretical results are Theorem~\ref{thm:ww:sob} and Theorem~\ref{thm:wsw:sob}. They state the required relations between the rate of writes $\lambda$ and the system size $N$. The relations can be affected by the ratio $\mu$ of used storage space, at which the system expansion is triggered. Setting $\mu$ to different values and assuming $b=1Gbps,v=16B$, we plot the relation curves for $\lambda$ and $N$ in Figure~\ref{fig:result:sob}. As expected, a smaller $\mu$ enables a system to process a heavier write workload. When its size is small, a system with a stable write workload can process writes in higher frequency than that with increasing writes. However, as the system size increases to a large number, e.g., 50, the influence of stable write workloads and that of linearly increasing writes do not differ much for the system.

\textbf{Bandwidth-Oriented Bounds.} We then consider the bandwidth-oriented bounds for stabilization with linearly increasing writes and with stable writes. The related theoretical results are Theorem~\ref{thm:ww:bob} and Theorem~\ref{thm:wsw:bob}. We plot the relation curves for the rate of writes $\lambda$ and the system size $N$ in Figure~\ref{fig:result:bob}. As compared to the results of storage-oriented bounds, the curves for the two write workloads are closer to each other, even when $N$ is small. For the same system size, one can obtain a higher upper bound on the rate of writes when the bandwidth-oriented bounds are employed. However, since a system must simultaneously satisfy both the storage and the bandwidth bounds, we must refer to the results of storage-oriented bounds in practice.

\textbf{Time-Oriented Bounds.} Now, we examine the results of the time-oriented bounds for clear stabilization with linearly increasing writes and with stable writes. The related theoretical results are Theorem~\ref{thm:nww:tob} and Theorem~\ref{thm:nwsw:tob}.  The relation curves for the rate of writes $\lambda$ and the system size $N$ are plotted in Figure~\ref{fig:result:tob}. Comparing the results of Figure~\ref{fig:result:tob} to the previous two figures, we can observe that the time-oriented bounds are much tighter. Therefore, a system employing clear stabilization can process a lower rate of writes than one that receives writes during its expansion process, although to stop receiving writes on stabilization can simplify the system design and implementation.

\subsection{Application Analysis}

Now, we examine the IoT use case presented in Section~\ref{sec:intro}. Before the administrator initiates a system expansion by one node, the rate of writes to the system is $4,800,000$. The average size of a written key-value is about $15*16=240$ bytes for the use case. Each node in the cluster has a network bandwidth of $1Gbps$. Based on the theoretical results of the paper, if we transform the relations between $\lambda$ and $N$ in the use case to the settings of Section~\ref{sec:numerical}, we have that $\lambda$ must be larger than \emph{approximately} $4,800,000$ for Figure~\ref{fig:result}. Accordingly, we can compute the corresponding $N$ for different stabilization choices. Note that the use case has stable writes on system expansion. Hence, if the data has taken up $\mu=0.5$ storage space of each node on average, the system can grow to about $13$ nodes (Figure~\ref{fig:result:sob}). If clear stabilization is used, the system can grow at most to about $17$ nodes (Figure~\ref{fig:result:tob}). These numbers are about the system size that the DHT-based Cassandra broke down at for the IoT use case.

To save the cluster from breaking down, two measures can be taken. First, based on the theoretical results, increasing the network bandwidth will raise the upper bounds for $\lambda$. Second, the system breaks down because of load rebalancing. Hence, if the system does not rebalance loads, the constraints will not take effect, although there would then occur the problem of partial overloading at some system nodes. A better choice is to update the DHT design for the IoT application scenarios. Since there is little room for improving the network bandwidth, the IoT data storage is in fact demanding a redesign of the scalable storage system architecture.

\section{Conclusion}

With regard to the new application scenarios of IoT, we study the scalability and the load balancing properties of DHT systems in this paper. Specifically, we address the load rebalancing problem of DHT, analyzing two factors that are rarely considered before. The two factors are the write workload on the system and the data movement during load rebalancing. It is the advent of IoT data, which come along with extremely heavy writes, that have changed the assumptions underlying the theoretical analysis of DHT systems. We formulate the load rebalancing problem for DHT, considering the impacts of write workloads and data movement during system expansion. For four application scenarios, we show that, the average network bandwidth of each node and the intensity of the average write workload are the two key factors that determine the feasibility of DHT load rebalancing. We prove theoretical bounds on the maximum rates of writes that a DHT-based system can handle, if the system is to rebalance loads in a node-by-node scale-out process. These results indicate that a redesign of the scalable storage system architecture is demanded for applications with heavy write workloads, e.g., IoT.

\balance

\bibliographystyle{IEEEtran}
\bibliography{IEEEabrv,ref}

\end{document}